\documentstyle[leqno]{article}

\begin{document}
\thispagestyle{empty}
\bibliographystyle{unsrt}

\begin{center}
{\large SCHR\"{O}DINGER CATS, QUANTUM SLINGS AND AZIMUTHAL EFFECTS IN
NUCLEUS-NUCLEUS COLLISIONS}\\

 I.~M.~DREMIN and V.~I.~MAN'KO\\

 {\it P.~N.~Lebedev Physical Institute, Leninskii Prospekt
53, Moscow 117924, Russia\\ E-mail: dremin@td.lpi.ac.ru} \\
\end{center}

\begin{abstract}
Confinement of a chromodynamical string can
result in specific effects in scattering processes and multiparticle production.
In particular, due to its rotation secondary fragments of the splitting apart 
string can move like stones thrown from a sling. That would be detected as the 
azimuthal asymmetry of particle distributions in individual events. Thus we 
propose to explain the elliptic flow observed in $AA$-collisions as a 
sling-effect. It can provide information about confinement of quarks inside 
particles or binding forces in nuclei. Nonclassical
states of the created particles like the Schr\"{o}dinger cats are produced.
Some classical and quantum-mechanical estimates
of possible effects are given. Experimental facts which can be used for their
verification are discussed.
\end{abstract}

The different types of nonclassical states of electromagnetic field quanta
attract a lot of attention in quantum 
optics~[1--4]
and in quantum mechanics~\cite{Nieto}. An essential property of the
nonclassical states of the field quanta is a difference of the particles'
statistics from the Poissonian statistics. Mechanisms of generating different
nonclassical states (squeezed states~\cite{Walls}, correlated 
states~\cite{Kurm,Bham}\,) for photons were discussed 
recently~[9--12].
The typical condition for nonclassical 
state generation is a sharp nonstationarity of the boundaries for photons or 
other quanta (e.g., in a cavity with moving walls). But an analogous
situation can take place in the process of high energy particle interactions,
as well. Thus, instant break-up of a particle or a nucleus and fast motion of some  
fragments can be considered as moving boundaries for other field quanta 
(nucleons, pions, gluons etc) involved in the process. In our opinion, such
collective effects as particle (or fragment) flows observed in high-energy 
nucleus reactions (momentum flow, directed flow, squeeze-out, elliptic flow)
are related to binding forces inside nuclei and to confinement of quarks in
particles. We will shortly discuss 
below some aspects of the approaches used in high energy physics and possible 
effects, which can be explained as obviously following from the simple 
``mechanical'' models.

In high energy physics, there coexist two seemingly incompatible approaches
to particle interactions. Those are the parton model, which has been proposed 
by Feynman and treats each particle as a conglomerate of a large number of 
point-like free partons, and the string model with quarks strongly confined 
inside hadrons. At the very beginning, these models did not compete because 
regions of their applicability were separated. The parton model pretended to 
describe the inelastic processes with high transferred momenta, while the 
string model was used for bound states. In quantum chromodynamics (QCD), 
quarks and gluons play the role of partons. Its property of the asymptotic 
freedom, i.e., of smallness of the coupling constant at small distances, has been 
used for the processes with high transferred momenta. For strings, one should 
consider the confinement phenomenon which can not be quantitatively described 
nowadays because it is purely nonperturbative effect.

However, step-by-step, the convergence of these models became quite clear. From 
one side, the higher perturbative approximations of QCD allowed us to describe
many properties of comparatively soft inelastic processes as multiplicity
distributions, inclusive rapidity distributions, some correlations, etc. From 
another side, the same properties have been described by the string 
(Lund) model with excitation and decay of strings considered as color dipoles.
Probably, it implies that above characteristics are not very sensitive to
some specific effects caused by the binding forces of quarks in such processes.

Our main goal is to search for such features which give
some information on confinement. We argue that the azimuthal collinearity
in individual events of particle (nuclear) interactions at low transferred
momenta could give some hints to confinement effects in inelastic processes.
Indeed, any string defines an axis in the space which, in general, differs
from the direction of the impinging particle. Thus there should be the azimuthal
asymmetry in a given event of a collision with the string. The strength of 
this effect should depend on string properties. In particular,
such asymmetry can be observed if the colliding objects (e.g. nuclei) start 
rotating after the
collision, in some sense keeping memory about their initial binding forces
during the process of their break-up.

Let us stress once more that only event-by-event analysis can provide
necessary information. From experimental point of view, it asks for
$2\pi $-geometry in azimuthal angles with constant acceptance.
Certainly, no azimuthal asymmetry exists when the trivial average over large 
ensemble of events is considered because there is no common spatial axis of
nucleus rotation or alignment of a string in different events. 
Besides, one should separate
the conservation of the transverse momentum from dynamical effects. But the
former effect must decrease at higher multiplicities. Two-jet production would be a
background of the effect sought for.

Here we give some examples borrowed from classical physics and nonrelativistic
quantum mechanics, which show how the binding forces can give rise to the
azimuthal collinearity in individual events.

Let us consider the classical nonrelativistic problem where the ball with the
radius $R$ and mass $M$ gets a blow with the momentum ${\bf p}$ at the impact
parameter $b$. The kinetic energy of the ball as a whole is 
$T_{\rm kin}=p^{2}/2M$ and its rotation energy is 
$T_{\rm r} =5p^{2}b^{2}/4MR^{2}$. The velocity at the
surface of the ball at the blow side is $v_{A}=p(1+5b/2R)/M$, and at the
opposite side $v_{B}=p(1-5b/2R)/M$. Thus $v_A$ is 3.5 times larger than
the velocity of the center of mass $v=p/M$ for $b=R$.
If a piece of mass $m_f$ of the surface
flyes off with the same velocity $v_A$, its momentum $p_{f}=p(1+5b/2R)m_{f}/M$
can be of the order of the initial momentum since the energy conservation asks
just for $(1+5b/2R)m_{f}/M \leq 1$. It is clear that this fragment moves in
an impact plane formed by the initial momentum vector and the center of the 
ball's mass. It reminds of a stone from a sling or of sparks from a firewheel 
or a grindstone. Therefore, there should be azimuthal inhomogeneity in an 
individual event. The similar effects are observed with a rod or a hard string.

The same arguments can be applied to nucleus collisions studied in Dubna and
Berkeley at moderate energies 2.5~GeV and 4.5~GeV per nucleon. For the sake of
simplicity, we consider only those events where the emulsion target is
hydrogen and work in the anti-laboratory system with the hydrogen nucleus
(proton) impinging on the heavy nucleus at rest and fragmenting it. Even though
the projectile is relativistic, the final system is not relativistic. The 
average rotation energy would be $\langle E_{\rm r} \rangle $=70 MeV at the primary 
energy 2.5~GeV, and 230~MeV at 4.5~GeV (with $\langle b^{2}/R^{2}\rangle 
$=0.5). For the isotropic decay of the nucleus in its rest system, the energy of 
each fragment  along a definite axis is $E_{0}/N$ and its rotation energy is
$\langle E_{\rm r} \rangle $ in the azimuthal plane. 
The azimuthal asymmetry is
large when they are of the same order, i.e., at $N\sim 20$. However, it can
be observable at lower number of fragments, as well. The effective value of the
azimuthal angle is estimated as $\tan \phi \vert _{\rm eff}\approx
(F_{f,\,y}/E_{f,\,x})^{1/2}\approx (1+3E_{\rm r}N/2E_{0})^{-1/2}$ 
that is equal (at $N$=4 and $E_{0}$=4.5~GeV) about 0.85 and differs from unity
for isotropic angles. One could try to ascribe to this effect the azimuthal 
collinearity observed in recent experiments~\cite{1}. This alignment has been 
seen in positive values of the second Fourier coefficient of the series 
expansion of particle distributions in differences of the azimuthal angles.

Let us consider the effects of the nucleus rotation in quantum mechanics.
This rotation is caused by bonds between different fragments inside a nucleus 
after one of the fragments is struck by a projectile moving along $z$-axis.
 The projectile feels these bonds as the fragment oscillations. We are 
interested in the angular distribution of scattered projectiles or of fragments
of the impinging nucleus. Then the
problem is similar to the old one treated by Bethe~\cite{2} for neutrons 
scattered by the paraffine molecule, but now at the other energy scale and 
other binding forces. The paraffine molecule reminds a string with very low 
binding energy. Now, for small polar angles, the angle between the transferred 
momentum and the direction of the strong bond ($x$ axis) is approximately equal 
to the azimuthal angle so that the distribution is (see~\cite{2}\,)
\begin{equation}
\frac {d\sigma }{d\cos \theta \,d\phi }\propto \exp [-2(1-\cos \theta )
(\epsilon_{1}\cos ^{2}\phi + \epsilon_{2}\sin ^{2}\phi )],    \label{1}
\end{equation}
where $\epsilon_{i}=E_{0}/h\omega _{i}$; \,$\omega _{i}$ being the oscillation
frequencies along the axes $x$ and $y$ with $\omega _{1}> \omega_{2}$. 
Integrating the polar angles, one gets at small difference between the 
frequencies $\Delta \omega =\omega _{1}-\omega _{2}$:
\begin{equation}
\frac {d\sigma }{d\phi }\propto (1+\frac {\Delta \omega }{\omega }\sin ^{2}\phi
)^{-1},     \label{2}
\end{equation}
i.e. the scattering is stronger in the impact $xz$-plane ($\phi =0$ or $\pi $ ) of
the strong bond and the projectile momentum. Note that the bond axis in each 
event has been chosen along the $x$-axis.
For the second Fourier coefficient describing the collinearity, one obtains
\begin{equation}
\langle \cos 2\phi \rangle = \frac {\pi }{4} \frac {\Delta \omega }{\omega },
\label{3}
\end{equation}
which is small but different from zero and qualitatively agrees with tendencies
observed in~\cite{1}. Namely, one can hope for smaller $\Delta \omega /\omega $
for heavier (and more symmetrical) nuclei, what is found in~\cite{1} as the
decrease of alignment with the atomic number increase.

One can calculate the azimuthal asymmetry in the scattering on a string in 
quantum mechanics in a following way. Consider the scattering of a particle on 
a system of two coupled scattering centers. If their recoil has been neglected,
then according to~\cite{br} (see also the book~\cite{gw}, Eq.~(311) in 
Chapter~11), the scattering amplitude is written as
\begin{eqnarray}
A^{(2)}&=&\left[A_{1}e^{-i{\bf qr}_{1}}+A_{2}e^{-i{\bf qr}_{2}}+A_{1}A_{2}
\frac {e^{ipR}}{R}\left(e^{i({\bf pr}_{1}-{\bf kr}_{2})}
+e^{i({\bf pr}_{2}-{\bf kr}_
{1})}\right)\right]\label{br1}\\
&&\times \left(1-A_{1}A_{2}\frac {e^{2ipR}}{R^2}\right)^{-1}.   
\nonumber
\end{eqnarray}
Here, $A_{i}$ are the scattering amplitudes on the centers $i$. 
 ${\bf r}_{i}$ denote the positions of the
centers, and we place one of them at the origin ${\bf r}_{1}=0$ and the second
at the distance $R$ along the $x$ axis, i.e., at ${\bf r}_{2}(R,0,0)$. 
Once again the direction of the strong bond is fixed.
The primary momentum along the $z$ axis is ${\bf p}(0,0,p)$, and the final 
momentum ${\bf k}$ is equal to the initial one in absolute value 
$\vert {\bf k}\vert $=$\vert {\bf p}\vert $. 
The transferred momentum is denoted by ${\bf q }={\bf k}-{\bf p}$. 
For two identical centers with opposite (color) charges, one gets $A_{1}=
-A_{2}=A$. For small amplitudes $A$ the multiple scattering effects can be
neglected and only two linear in $A$ terms of Eq.~(\ref{br1}) survive.
Then one obtains
\begin{equation}
A^{(2)}\approx A(1-e^{-ipR\sin \theta \cos \phi }).
\label{br2}
\end{equation}
 The scattering amplitude $A^{(2)}$ vanishes at $R\rightarrow 0$ due to the  
color screening (the so-called color transparency effect). 
The differential cross section is 
\begin{equation}
\frac {d\sigma }{d \Omega }=\vert A^{(2)}\vert ^{2}=2\vert A\vert ^{2}
[1-\cos (pR\sin \theta \cos \phi )].    \label{br3}
\end{equation}
The cross section vanishes for $\cos \phi =0 (\phi =\pm \pi/2)$ i.e. in plane
perpendicular to the impact plane that corresponds to elliptic flow.

We have argued that different type azimuthal asymmetry can appear.
The transition to ever higher energies, when the inelastic interactions of
particles are studied, asks for the relativistic treatment of strings which
is not yet developed. Here, one can hope that the analogy from the low energy
region would work with some ``natural'' modifications, and one can propose such
a picture where the string is treated as quark oscillations inside some
confining potential and the external blows give rise to excitations and
breaking of strings. The transverse momentum conservation would lead to
azimuthal asymmetry of fragments.

Such breaks could be described as abrupt vanishing
of the potential that would produce the quantum sling-effect suggested by Hacyan
in~\cite{3}. He has solved the Schr\"{o}dinger equation for the parametric 
oscillator with potential becoming zero at some time what describes a quantum
sling. The specifics of the quantum sling is that the system
of the fragments produced is a nonclassical state of even and odd coherent
states introduced in~\cite{dmm} and modelling the Schr\"{o}dinger cat states.
They have been seen for the photons in cavities~\cite{h}, and never discussed
in particle or nuclear physics. The quantum-sling mechanism considered here
could be in charge of some states like the Schr\"{o}dinger cat in particle
production. Its typical feature is the dominance of either even, or odd
number of produced field quanta in a single process. Let us note that the 
isospin conservation could mimic such states as well.

There exists the special axis of a string and, therefore, the azimuthal 
asymmetry
should be observable in individual events. It is preferred to work with the
non-relativistic objects to treat the string rotation. Let us estimate
what restrictions it would impose for scattering of leptons on hadrons (or
nuclei). The simplest situation is where the target is almost unexcited but
rotates as a whole. Then the nonrelativistic condition is provided by
\begin{equation}
E_{r}=W-M\approx M(1+\frac {1-x}{x}\frac {Q^2}{M^2})^{1/2}-M\approx \frac {1-x}
{2x}\frac {Q^2}{M}\ll M  \label{4}
\end{equation}
\noindent  and leads to
\begin{equation}
Q^2 \ll \frac {2x}{1-x} M^2.       \label{5}
\end{equation}
Here $W$ is the total target energy in its rest system, $Q^2$ is the squared
transferred four-momentum (photon virtuality), $x$ is Bjorken variable. One
concludes that the nonrelativistic region is limited by very small values
of $Q^2$, but heavier targets widen it.

There are many strings in typical inelastic processes at high energies, and
the background due to their different orientations in space can be strong.
However, one hopes that in some rare events a single string (or Pomeron?)
can play a dominant role giving rise to strong alignment. It would ask for
high orbital moments in such collisions. The events
of that kind were, probably, seen in cosmic rays~\cite{4} if their
alignment survives the secondary interactions in the atmosphere. To prove it
one needs Monte-Carlo calculations.

At the same time, one can envisage that the ratio $\Delta \omega /\omega $
in Eqs (\ref{2}), and (\ref{3})
becomes large due to strong disbalance of forces along the string and
perpendicular to it. It would enlarge the azimuthal asymmetry.

Thus, we propose to experimentalists to look for the azimuthal collinearity in
inelastic events by evaluating the alignment coefficients. They are introduced
in~\cite{1} for an individual event with multiplicity $n$ as
\begin{equation}
\beta _{2}=\frac {\sum _{j}\sum _{i>j} \cos 2(\phi _{i}-\phi _{j})}
{\sqrt {n(n-1)}} , \label{be}
\end{equation}
where $\phi _{i}-\phi _{j}$ denotes the difference between the azimuthal 
angles of particles $i$ and $j$. The average value of $\beta _2$ over an ensemble 
of events must be calculated. An excess over the background of statistical 
fluctuations provided by traditional models, with jets accounted, 
would be a signature of confinement. Our preliminary analysis of some Pb-Pb
collisions at 158 GeV supports the optimistic approach to the problem.
For analysis of "Mercedes"-type events one should use $\beta _3$ instead $\beta _2$
with a factor in cosine being 3 instead 2. In general, $\beta _m$ with any
integer $m$ can be used to study "$m$-leaves" topology of events.
Another signature of creating the field quanta in nonclassical states is the 
difference of the multiplicity distributions from the Poissonian statistics
and an excess of even or odd states.
Beside Fourier coefficients, one can use the wavelet analysis~\cite{5} or other
correlation methods~\cite{6} for event-by-event studies. We hope that these
methods will help separate the nuclei rotations and the string tension to get
some confinement parameters. We will study these problems in more detail later.

The work was partially supported by the Russian Federation Ministry for
Science and Technology, by Russian fund for basic research (grant 96-02-16210)
and by INTAS (grant 93-79).

\end{document}